\documentclass[10pt]{article}
\usepackage[utf8]{inputenc}
\usepackage{amsmath}
\usepackage{graphicx,tabularx}
\usepackage{amsfonts,amsthm,eucal,amssymb,amscd}
\usepackage{color}
\usepackage[all]{xy}
\usepackage{epstopdf}
\usepackage{upgreek}
\usepackage{bbold} 
\usepackage{glossaries}
\usepackage{tikz-cd}

\begin{document}

\newtheorem{theo}{Theorem}[section]
\newtheorem{definition}[theo]{Definition}
\newtheorem{lem}[theo]{Lemma}
\newtheorem{prop}[theo]{Proposition}
\newtheorem{coro}[theo]{Corollary}
\newtheorem{exam}[theo]{Example}
\newtheorem{rema}[theo]{Remark}
\newtheorem{example}[theo]{Example}
\newcommand{\ninv}{\mathord{\sim}} 
\newtheorem{axiom}[theo]{Axiom}

\title{Indistinguishability and the origins of contextuality in physics}

\author{
J. Acacio de Barros$^{1}$, Federico Holik$^{2}$ and D\'{e}cio
Krause$^{3}$}

\maketitle

\begin{center}

\begin{small}
1- School of Humanities and Liberal Studies, San Francisco State University, San Francisco, CA, USA\\
2- Institute of Physics La Plata (IFLP), Buenos Aires, Argentina\\
3- Department of Philosophy, Federal University of Brazil in Santa
Catarina, Florian\'opolis, SC, Brazil.
\end{small}
\end{center}

\vspace{1cm}

\bigskip
\noindent

\begin{small}
\centerline{\em Key words: contextuality, quasi-sets, quantum
indistinguishability}
\end{small}

\begin{abstract}
In this work we discuss a formal way of dealing with properties of
contextual systems. Our approach is to assume that properties
describing the same physical quantity, but belonging to different
measurement contexts, are indistinguishable in a strong sense. To
construct the formal theoretical structure, we develop a description
using quasi-set theory, which is a set-theoretical framework built
to describe collections of elements that violate Leibnitz's
principle of identity of indiscernibles. This allows us to consider
a new ontology in order to study properties of quantum systems.
\end{abstract}

\section{Introduction}

The concept of a property of a quantum system is hard to define
consistently. For instance, in a famous paper
\cite{kochen_problem_1967}, Kochen and Specker (KS) showed that
attempting to assign truth values to a quantum property
\cite{von_neumann_mathematical_1983,piron_foundations_1976}, as
predicted by the algebra of observables in a Hilbert space, may
result in logical contradictions unless we assume that properties
depend on which other properties are being simultaneously observed
with it. This dependency is what is known in the literature as
contextuality, reflecting the idea that properties are
context-dependent ---context being defined by the other
simultaneously observed quantities. Quantum contextuality creates
the difficulty that the value of a property becomes dependent on the
observer's choice of what context to measure it in, i.e. with which
other properties, and not with an intrinsic characteristic of the
measured quantum system.

The contextual dependency of properties of quantum particles seems
to be an essential aspect of the microscopic world. It is related
not only to the Kochen-Specker apparent paradox above mentioned, but
also to Bell's proof of the incompatibility of quantum predictions
with local-realism \cite{bell_einstein-podolsky-rosen_1964}. It is
also an important resource in quantum computation
\cite{veitch_negative_2012}, and perhaps the main reason why quantum
computers outperform their classical counterparts
\cite{howard_contextuality_2014}. Thus, it is not surprising that
extensive research in contextuality has happened in recent decades,
with entire conferences devoted to it (such as the Quantum
Contextuality in Quantum Mechanics and Beyond workshop in Prague, or
the Winer Memorial Lectures at Purdue University).

This paper examines contextuality in physics by extending a point of
view put forth on references  \cite{de_barros_contextuality_2017}
and \cite{kurzynski_contextuality_2017}, namely that quantum
indistinguishability is connected to contextuality. In
\cite{de_barros_contextuality_2017}, we argued that the
indistinguishability of particles, expressed mathematically by a
set-theoretical construct where the law of identity of
indiscernibles is violated, invalidates the contradiction argument
put forth by Kochen and Specker (KS) in \cite{kochen_problem_1967}
(see Section \ref{sec:describing-quantum-systems} for a sketch of
the KS argument). In their paper, KS discuss the concept of quantum
properties, represented by self-adjoint operators in a Hilbert space
\cite{von_neumann_mathematical_1996}, and show that attempts to
assign truth values to those properties in different experimental
contexts fail if we assume that such properties are
context-independent. We argued that KS argument was not necessarily
valid for quantum systems because, since particles are in principle
indistinguishable, it is not possible to say that we are talking
about  property X or Y of particle A or particle B. All we know is
that, in the case of A and B, we have two particles, and that they
have different properties X and Y, but we cannot, \textit{in
principle}, know which one has property X or property Y.

Here we extend the above notion to show that it is not just the
indistinguishability of particles that may be at play in physics,
but also the indistinguishability of properties, seem here as a
representation of the conjunction of a specific measurement
apparatus and a physical (quantum) system (e.g. a particle). In
other words, as in the case of particles, we cannot know which
property we are talking about. Two properties, A and A', may be
indistinguishable, and this indistinguishability leads to apparent
paradoxes if we treat them, as often is done in physics, as one and
the same.

The idea of indistinguishability of properties can be seen as
stemming from \cite{dzhafarov_contextuality_2014}. In that
reference, it is argued that contextuality is about the
\emph{identity} of properties, in the sense that, properties taken
from different contexts must be considered different. In reference
\cite{dzhafarov_contextuality_2014}'s approach, properties are
assumed to obey the classical theory of identity, formalized in it
by the use of random variables in a probability space. The
connection between identity and contextuality seems to  also appear
in the formalism of QM. Take the case of three observables, such as
Bell's case where observables $A$ and $A'$ refer to Alice's
observations and $B$ to Bob's. The self-adjoint operators in the
subspace representing Bob's observable $B$ are the same when Alice
measures $A$ or $A'$, regardless of whether Alice's choice of
measurement. It seems to be the same property, among the different
contexts of measurement. The experimental setup is the same as well:
there is an operationally identifiable procedure that allows us to
say that we are measuring $A$, on each instance. But yet, if we
assume that all those instances represent one and the same property,
we are lead to contradictions. This situation leads the authors of
\cite{dzhafarov_contextuality_2014} to conclude that the a
properties associated to the same physical quantity, but considered
in different contexts, cannot be the same. So, the distinction
between $B$ in context $A$ and $B$ in context $B'$ is only that they
are different properties, but not distinguishable. If they were
distinguishable,  i.e. if Bob could run an experiment where he
could, just looking at $B$, determine Alice's choice of measuring
$A$ or $A'$, they would be able to use this to signal to each other
in a superluminal way. This, of course, is forbidden by quantum
mechanics \cite{dieks_communication_1982}.

How to make sense of the assertion that properties are different
among different contexts, while it is, at the same time, the same
quantity being measured? Reference
\cite{dzhafarov_contextuality_2014} takes the approach of just using
different random variables for each context. Though this is
logically consistent, and probably work for the general case, here
we take an alternative stance: we will assume that properties
---representing the same quantity to be measured--- are
indistinguishable among the different contexts, but yet, not the
same. To do so, we use quasi-set theory \cite{krause_formal_1995},
which is a logical formalism developed to deal with collections of
truly indistinguishable entities. Quasi-sets have been applied to
quantum mechanics before, in order to describe quantum
non-individuality
\cite{domenech_discussion_2007,domenech_q-spaces_2008,domenech_quantum_2010},
and as above mentioned, to avoid Kochen-Specker-type contradictions
\cite{de_barros_contextuality_2017}. We see our approach as having
two advantages over the more general one of
\cite{dzhafarov_contextuality_2014}. First, it comes from a strong
ontological assumption that quantum systems are truly, and in
principle, non-distinguishable. This does not seem to be an
epistemic issue, as there are real consequences for this
indistinguishability at the level of particles (i.e. different
quantum statistics, or other quantum effects such as Bose-Einstein
condensates). The second advantage is that our approach stays closer
to the Hilbert space representations of quantum theory by using the
same random objects for all contexts.

This paper is organized as follows. For each section, we try to
provide an intuitive and simplified version of the concepts
discussed, and then present them in a more formal way. Our goal is
to make this paper more accessible to a broader audience who may not
be familiar with some of the ideas used here. In Section
\ref{sec:describing-quantum-systems} we discuss the concept of
properties for quantum systems, and show how they are problematic
because of contextuality. In Section
\ref{sec:indistinguishability-particles} we show how we can
represent indistinguishability of particles using the formalism of
quasi-sets, constructed formally as an axiomatic theory where
Leibniz's principle of identity is violated. Then, in Section
\ref{sec:indistinguishability-properties} we extend the ideas of
Section \ref{sec:indistinguishability-particles} to properties, and
we show how when properties are indistinguishable the usual
contextual inequalities are not derivable. We end the paper with
some conclusions and final remarks in Section \ref{sec:conclusions}.

\section{Describing quantum systems}\label{sec:describing-quantum-systems}

In order to discuss the quantum case, let us start with the general
concept of properties\footnote{We shall not give a detailed theory
of properties relevant to the empirical sciences, and the interested
reader may find references such as
\cite{krantz_foundations_1971,krantz_foundations_1989},
\cite{luce_foundations_1990} or \cite{suppes_representation_2002} as
useful and comprehensive resources; see also
\cite{birkhoff_logic_1936,jauch_foundations_1968,foulis_half-century_1999}
for the notion of property in the quantum logical approach, which
plays a key role in quantum mechanics and the derivation of the
Kochen-Specker theorem.}. Intuitively, a property is a
characteristic or quality of something. For example, when we say
that ``the sky is blue'', the color blue is a characteristic of how
we perceive the sky. In physics, when we talk about properties of a
system, we mean something similar: what characteristics this system
has. For instance, if we say that a metal block is 32cm long or 400g
in mass, these statements represents properties the block has: the
property of being 32cm long and having a mass of 400g. This concept
of property is straightforward in classical physics, where we can
talk about volumes of solids, temperature of an object, or energy of
a system, to mention a few.

The most basic type of property is a binary property, i.e. a quality
of the system that is either true or false. In other words, we can
probe whether a system has or does not have the property. Some
simple examples of binary properties are represented in the
following questions: ``Does Federico have the property of being
tall?'' or ``Is it cold today?'' Of course, such casual properties
are not what we are talking about in physics, and we need to be more
precise, going beyond defining which height we think is the minimum
for being considered tall, or what temperature below which we feel
it is cold.

As such, in physics we need to talk about more complicated
properties that can account for more specificity, such as today's
temperature in C (it is 13 C outside, reads someone's thermometer)
instead of simply saying it is cold or not. It is straightforward to
see that even such properties are made up of several binary
properties. Let us examine the temperature example. To measure the
temperature means to give a number that is within the range of the
thermometer (say -20 to 60 C) and that is consistent with its
precision ($\pm 0.5$ C). Consider the following series of statements
that can be either true or false. $A_{-20}$=``the tip of the mercury
column of the thermometer is in the interval $-20.0\pm 0.5$'',
$A_{-19}$=``the tip of the mercury column of the thermometer is in
the interval $-19.0\pm 0.5$'', $A_{-18}$=``the tip of the mercury
column of the thermometer is in the interval $-18.0\pm 0.5$'', and
so on until $A_{60}$=``the tip of the mercury column of the
thermometer is in the interval $60.0\pm 0.5$''. Each of those
statements are compatible (i.e. one can verify the veracity of each
of those questions simultaneously), and they are complementary (i.e.
one and only one of them may be true at a given time). So, the
statement ``it is 13 C outside'' means that the proposition
$A_{13}$=``the tip of the mercury column of the thermometer is in
the interval $13.0\pm 0.5$'' is true, whereas all other
complementary propositions $A_i$, $i\neq 13$, are false. In fact,
any numerical property could be represented this way, as made up of
several individual and compatible binary properties. So, properties
such as mass, charge, position, temperature, entropy, length, etc.
can be though as a combination of a large (sometimes infinite)
number of binary properties where only one of them can be true at a
time.

There is an interesting connection between properties and ontology
in classical physics. For example, in classical Newtonian physics,
physical systems are composed of  particles, whose fundamental
properties are their mass, position, and velocity. Because mass is a
constant for Newtonian particles, only position and velocity can
vary, and the value of all particles' positions and momenta are
called the  \textit{state of the system}. Any other property, such
as the system's energy, temperature, or length (if talking about a
solid made of particles itself) are definable in terms of the
properties of the fundamental constituents of the system, the
particles, namely their position and velocity. Furthermore, given a
system of particles and their interactions, knowing their state,
i.e. their position and velocity, completely determines their future
state, and therefore any properties associated to the system. But,
more importantly, given that properties are definable in terms of
two quantities that are simultaneously measurable, position and
velocity, it follows that each property  can be though as a subset
of the space of all possible positions and velocities (usually
$\mathbb{R}^{6N}$, where $N$ is the number of particles and $6$ the
number of components of the position and velocity vectors necessary
to describe the particle). Therefore, if we wish to define an
algebra of properties, this algebra would be simply a Borel algebra
on $\mathbb{R}^{6N}$.

In quantum theory, things are very different. First, there is no
simple and widely accepted ontology for quantum systems similar to
the classical one \footnote{It is important to mention that Bohm's
theory \cite{bohm_suggested_1952-1} provides an ontology close to a
classical one, but it is far from being widely accepted among
physicists \cite{schlosshauer_snapshot_2013}. One of the main
reasons for this, is, perhaps, that its hidden-variables behave in a
manifestly non-local way. Furthermore, the hidden-variables
introduced are not of much use in practice, given that they cannot
be manipulated in the lab (due to its hidden character). Thus, they
play only an ad-hoc explanatory role, without giving place to any
relevant predictions. Our approach in this work aims to stay closer
to most physicists guiding intuitions in their practice.}. Second,
the state of a quantum system is not definable in terms of the
position and velocity of its particles. The reason is a fundamental
one: contrary to classical particles, position and velocity of a
quantum particle cannot be, in principle, measured simultaneously
with as much precision as we wish. Therefore, the idea of defining
properties as subsets of  $\mathbb{R}^{6N}$ is not a straightforward
matter \footnote{Attempts to do so lead to quasi-probability
distributions \cite{wigner_quantum_1932}.}.

Instead, binary properties are represented in quantum theory by
projection operators in a separable Hilbert space, $\mathcal{H}$
\cite{birkhoff_logic_1936,domenech_quantum_2010}. The Hilbert space
itself is determined by the number of such binary properties that we
can maximally observables. More complex outcomes of experiments and
their associated properties are modeled by self-adjoint operators in
$\mathcal{H}$. Due to the spectral decomposition theorem, Hermitian
operators can always be written as sums of projection operators. In
other words, we can think of Hermitian operators as made up of
several binary properties, which, due to their connections to
experiments, are called \textit{observables}. Thus, observables and
the Hilbert space are dependent not only on the system, but on our
ability to extract information from this system. The more
information, the larger the Hilbert space becomes.

To give a less abstract example, take the case of a single electron.
If we were only able to measure its position on the $x$ direction,
its Hilbert space would be $\mathcal{L}^2$, the space of all square
integrable functions, and a vector in this space would be a function
$\psi \in \mathcal{L}^2$ whose absolute value squared,
$|\psi(x)|^2$, at $x$ gives the probability density that the
electron is found between $x$ and $x+dx$ if a measurement of
position is performed. In this Hilbert space, the position operator
is simply $x$. Correspondingly, the observable associated to the
property ``momentum'' is the operator $i\hbar \partial /\partial x$
on $\mathcal{L}^2$. However, electrons have another property of
interest:  spin. The Hilbert space for spin 1/2, as is the case for
the electron, is $\mathbb{C}^2$. So, if we were to only measure
position or momentum, the Hilbert space would be $\mathcal{L}^2$; if
we were to only measure spin (regardless of direction), the Hilbert
space would be $\mathbb{C}^2$; if we were to measure both position
and spin, the Hilbert space would be $\mathcal{L}^2\otimes
\mathbb{C}^2$. Things get more complicated as we increase the number
of particles, or if we increase the number of possible observables.

As mentioned above, properties related to a Hilbert space through
observable operators. Given a quantum system $S$, we can construct a
Hilbert space $\mathcal{H}$, whose basis represent a set of possible
projection operators that completely span $\mathcal{H}$. This set of
projection operators provide a maximal set of compatible properties
of $S$. We can then use those properties to create more complex
properties in the form of Hermitian operators.

Again, let us explore this with a simple example: a
three-dimensional Hilbert space, $\mathbb{C}^3$. Since this space is
three-dimensional, it follows that a basis for this space is
constituted of three linearly independent vectors, say
\[
\mathbf{e}_{1}=\left(\begin{array}{c}
1\\
0\\
0
\end{array}\right), \,\,\,\,\,\mathbf{e}_{2}=\left(\begin{array}{c}
0\\
1\\
0
\end{array}\right), \,\,\,\,\,\mathbf{e}_{3}=\left(\begin{array}{c}
0\\
0\\
1
\end{array}\right).
\]
and the corresponding projectors associated to each vectors are
\[
P_{\mathbf{e}_{1}}=\left(\begin{array}{ccc}
1 & 0 & 0\\
0 & 0 & 0 \\
0 & 0 & 0
\end{array}\right),
\,\,\,\,\, P_{\mathbf{e}_{2}}=\left(\begin{array}{ccc}
0 & 0 & 0\\
0 & 1 & 0 \\
0 & 0 & 0
\end{array}\right),
\,\,\,\,\, P_{\mathbf{e}_{3}}=\left(\begin{array}{ccc}
0 & 0 & 0\\
0 & 0 & 0 \\
0 & 0 & 1
\end{array}\right).
\]
We can see that
$P_{\mathbf{e}_{1}}+P_{\mathbf{e}_{2}}+P_{\mathbf{e}_{3}}=\hat{1}$,
where $\hat{1}$ is the identity matrix, which is a consequence of
$\mathbf{e}_{1}$, $\mathbf{e}_{2}$, and $\mathbf{e}_{3}$ forming a
basis for $\mathcal{H}$. It is easy to create now, in this
formalism, an observable that is associated to the property of
having values $1$, $2$, and $3$ as simply
$P_{\mathbf{e}_{1}}+2P_{\mathbf{e}_{2}}+3P_{\mathbf{e}_{3}}$.

Once you have a basis for $\mathcal{H}$, you can define observables
as above. However, it is always possible to define another basis. In
quantum theory, this new basis will correspond to new observable
properties. What is important here is that a property with a
definite value in one basis may not be associated with a state that
has definite values for another property (represented by another
basis). Furthermore, once we measure the observable associated to
the new basis, and find out a value for a given property, the new
state of the system will be associated to this property, and the old
basis (and their corresponding properties) will not have definite
values anymore. An observation (or measurement) affects the state of
the system.

In other words, the sequential observation of properties in
different basis may lead to changes in the outcomes of past
observations. Properties become dependent on the how we observe
them: if we first observe A and then B, we may get something
different from observing B and then A. Even more importantly, if we
observe A, B, and then A, the second time we observe A its value may
be different. Additionally, attempts to assign values to properties
that are independent of how we observe them will lead to
inconsistencies.

The above argument is at the core of KS's theorem
\cite{kochen_problem_1967}. In an example provided by Cabello et al.
\cite{cabello_bell-kochen-specker_1996}, we start with a specific
set of projection operators $P_i$, $i=1,\ldots,18$, in a Hilbert
space of dimension four. The set $\{ P_i \}$ is selected such that
there are 9 contexts such that the sum of the four $P_i$'s in it is
the identity operator (e.g. $P_1+P_2+P_3+P_4=\hat{1}$). Furthermore,
the contexts are selected such that each projector appears twice,
once in two different contexts (e.g. $P_1+P_2+P_3+P_4=\hat{1}$ and
$P_1+P_5+P_6+P_7=\hat{1}$). The consequence is that we have 9
equations, one for each context, where they all add to $\hat{1}$,
and such that each operator appears twice. Now, the contradiction
comes from the fact that if we associate to each projector a
property of being 1 (for true) and 0 (for false) as the same in all
contexts, the sum of all projectors' values on the 9 equations adds
to an even number (each appears twice). However, since each equation
adds to one, their sum needs to add to nine, which is odd, and we
reach a contradiction. This is the essence of the KS theorem: the
assumption that the values of $P_i$ are independent of context lead
to a contradiction.

One may object that the above arguments were focused solely on cases
where the property of the system is known with certainty, i.e. one
can assign to it a truth value. One may argue that in quantum
physics, properties are not deterministic, and one must talk about
probabilities, which would make the above arguments not apply.
However, this is not the case. It is possible to show that the
underlying assumption that properties have a value, even though we
may not know what they are and represent them with a probability
function, is incompatible with quantum theory
\cite{fine_joint_1982}. The reason is that standard probability
theory assumes an underlying consistency through a Boolean algebra.

To accommodate the quantum predictions, one must either expand the
number of properties to include other properties that are
co-measured (see \cite{dzhafarov_contextuality-by-default_2017} and
references therein), or one needs to modify probability theory,
either by allowing probabilities to take negative values
\cite{spekkens_negativity_2008,de_barros_negative_2014,de_barros_negative_2016},
by changing the rule for adding probabilities
\cite{suppes_existence_1991,de_barros_probabilistic_2010}, by
modifying the algebra of events
\cite{foulis_half-century_1999,narens_probabilistic_2014,holik_discussion_2014},
or by rethinking about measurement outputs as depending on all
components of an examined experiment
\cite{khrennikov_contextual_2009}. But regardless of how we choose
to deal with such issues, the key aspect of the quantum world is
that quantum properties are not definable in a consistent way if we
require classical logic and context independence. This quantum
contextuality is essential for any ontology associated to it, and we
will explore it in more details in  Section
\ref{sec:indistinguishability-properties}.

\section{Indistinguishability of particles}\label{sec:indistinguishability-particles}

There are two remarkable features that characterize compound quantum
systems. One of them is \emph{entanglement}, that can be interpreted
as the impossibility of describing certain quantum correlations by
appealing to mixtures of classical correlations
\cite{werner_quantum_1989}. The other feature ---the one that we are
interested in--- is indistinguishability: when quantum systems of
the same kind are put together, they display statistics which are
very different from those which are used to describe distinguishable
entities\footnote{We emphasize that the indistinguishability of
particles is an ontological assumption. For example, as mentioned
before, in Bohmian theory
\cite{bohm_suggested_1952-1,bohm_suggested_1952} the ontology is
classical, with quantum effects originating from a quantum
potential. The anti-symmetrization or symmetrization of the wave
function, thus, in this theory, leads to non-local effects on the
particles due to the quantum potential. Though the authors of this
paper are sympathetic to Bohm's approach, we are also aware that the
majority of the physics community rejects it, perhaps mainly because
of its classical ontology. In this paper, we embrace the quantum
weirdness, and try to explain quantum effects based on a
non-classical ontology of indistinguishable particles. }. This is
expressed in the symmetrization postulate and the celebrated Bose
and Fermion statistics. This feature lies behind very important
fields of research, such as the study of Bose-Einstein condensates.
In recent years, the difference between entanglement and
indistinguishability has been studied in detail: it turns out that
these are very different physical features, in the sense that a
quantum system can be prepared in a fully symmetrized state in which
no non-local correlations are present
\cite{plastino_separability_2009}. This distinction leads us to the
question of whether it is possible to consider quantum
indistinguishability as a resource. What is the relationship, if
any, between quantum indistinguishability and contextuality? In
order to explore possible answers for these questions in the
following sections, let us first review the formalism for
indistinguishable particles.

Let us illustrate how the formalism works for only two fermions. For
this case, if we know that one particle is in state $|b\rangle$ and
the other in state $|a\rangle$, then, using the symmetrization
postulate, the joint state will be given by
\begin{equation}\label{indistinguishable}
|\psi\rangle=\frac{1}{\sqrt{2}}(|a\rangle\otimes|b\rangle-|b\rangle\otimes|a\rangle)
\end{equation}
The above state means that there is one particle in each state. But
the symmetrization tells us that we cannot tell which one is which:
a permutation of the particles yields an overall minus sign, and
thus, the probabilistic predictions of the theory are exactly the
same. This situation lead many authors ---including E.
Schr\"{o}dinger--- to conclude that quantum systems, in certain
situations, cannot be considered as individuals
\cite{krause_quasi-set_1999}.

In the position representation, the wave function associated to our
two-particle state in (\ref{indistinguishable}) is given by
\begin{equation}
\psi(x,y)=\frac{1}{\sqrt{2}}(\psi_{a}(x)\otimes\psi_{b}(y)-\psi_{b}(x)\otimes\psi_{a}(y)),
\end{equation}
where $x$ and $y$  are the coordinates of particles 1 and 2. When
$x\longrightarrow y$, i.e. when the particles are close to each
other, we observe that the square modulus of the wave function
\begin{equation}
|\psi(x,y)|^{2}=\frac{1}{2}(|\psi_{a}(x)\psi_{b}(y)|^{2}+|\psi_{b}(x)\psi_{a}(y)|^{2}-2\Re(\psi_{a}(x)\psi_{b}(y)\psi_{b}^{\star}(x)\psi_{a}^{\star}(y)))
\end{equation}
tends to zero. This implies that no two fermions can be found
occupying the same state. If the wave functions $|\psi_{a}(x)|$ and
$|\psi_{b}(y)|^{2}$ have compact support, we see that there are no
indistinguishability effects when $|x-y|$ is big enough. Thus, when
the particles get close each other, we can go continuously from a
distinguishability for all practical purposes situation, to a
non-distinguishability one.

The state of a Bosonic system is similar to
(\ref{indistinguishable}), but there is ``$+$" sign instead of a
``$-$". In the first case, the states are symmetric under
permutation of particles, while in the second, anti-symmetric. This
changes the statistics considerably: unlike Fermions, it is possible
to have an arbitrary number of Bosons occupying the same state.

In order to study the case with arbitrary particle number, it is
useful to consider the Fock-space formalism. The standard Fock-space
is built up from the one particle Hilbert spaces as follows. Let
$\mathcal{H}$ be a Hilbert space and define
\begin{eqnarray}
&{\mathcal{H}}^{0}&= {\mathbb{C}},
\nonumber\\
&{\mathcal{H}}^{1}&= {\mathcal{H}},
\nonumber\\
&{\mathcal{H}}^{2}&= {\mathcal{H}}\otimes{\mathcal{H}},
\nonumber\\
&\vdots&\nonumber\\
&{\mathcal{H}}^{n}&= {\mathcal{H}}\otimes \cdots \otimes
{\mathcal{H}}.
\end{eqnarray}
The Fock-space is thus constructed as the direct sum of $n$ particle
Hilbert spaces,\begin{equation} {\mathcal{F}}=
\bigoplus^{\infty}_{n=0} {\mathcal{H}}^{n}.
\end{equation}
When dealing with bosons or fermions, the symmetrization postulate
must be imposed. Thus, given a vector $v=v_{1}\otimes\cdots\otimes
v_{n}\in {\mathcal{H}}^{n}$, define
\begin{equation}
\sigma^{n}(v)=(\frac{1}{n!})\sum_{P}P(v_{1} \otimes\cdots\otimes
v_{n})
\end{equation}
and
\begin{equation}
\tau^{n}(v)=(\frac{1}{n!})\sum_{P}s^{p}P(v_{1} \otimes\cdots\otimes
v_{n}),
\end{equation}
where
\[ s^{p} = \left\lbrace
  \begin{array}{c l}
    1 & \text{if p is even},\\
    -1  & \text{if p is odd}.
  \end{array}
\right. \] Let
\begin{equation}
{\mathcal{H}}^{n}_{\sigma}= \{\sigma^{n}(v):v\in {\mathcal{H}}^{n}
\}
\end{equation}
and
\begin{equation}
{\mathcal{H}}^{n}_{\tau}= \{\tau^{n}(v):v\in {\mathcal{H}}^{n} \}.
\end{equation}
Thus, we have the Fock-space
\begin{equation}
{\mathcal{F}}_{\sigma}= \bigoplus _{n=0}^{\infty}
{\mathcal{H}}^{n}_{\sigma}
\end{equation}
for bosons and
\begin{equation}
{\mathcal{F}}_{\tau}= \bigoplus^{\infty}_{n=0}
{\mathcal{H}}^{n}_{\tau}
\end{equation}
for fermions.

The standard wave mechanics approach to the description of
multi-particle systems uses the kinetic energy operator
\begin{equation}
T_{n} =  \sum_{i=1}^{n}T_{1}(r_{i})
\end{equation}
for $n$ particles, where $T_{1}(r)=
-(\frac{\hbar^{2}\nabla^{2}}{2m})$. A similar expression holds for
the external potential. For a pairwise interaction potential, we
have
\begin{equation}
V_{n}= \sum_{i>j=1}^{n}V_{2}(\mathbf{r}_{i},\mathbf{r}_{j }).
\end{equation}
The total Hamiltonian operator is thus given by
\begin{equation}\label{e:nhamiltonian}
H_{n}=\sum_{i=1}^{n}[(-\frac{\hbar^{2}\nabla_{i}^{
2}}{2m})+V_{1}(\mathbf{r}_{i} )+\sum_{i>j=1}^{n}
V_{2}(\mathbf{r}_{i},\mathbf{r}_{j})].
\end{equation}
The $n$-particles wave function can be expressed as
\begin{equation}
\Psi_{n}(\mathbf{r}_{1},\ldots,\mathbf{r}_{n},t),
\end{equation}
and it is a solution of the Schr\"{o}dinger's equation
\begin{equation}\label{e:SchrodingerEquation}
H_{n}\Psi_{n}=i\hbar\frac{\partial}{\partial t}\Psi_{n}.
\end{equation}

\section{Indistinguishability of properties}\label{sec:indistinguishability-properties}

We now move to discuss what we call the indistinguishability of
properties, starting with the concept of properties. Suppose that we
have the quantum system formed by parts $S_{1}$ and $S_{2}$. The
observables associated to the compound system will be represented by
the algebra $\mathcal{B}(\mathcal{H})$, with
$\mathcal{H}=\mathcal{H}_{1}\otimes\mathcal{H}_{2}$. For any
observable $A$ of $S_{1}$, we may consider the observable of the
compound system $A\otimes B$. The interpretation of $A\otimes B$ is
that we measure $A$ in $S_{1}$ and $B$ in $S_{2}$. But for each
$A\in\mathcal{B}(\mathcal{H}_{1})$, there are infinitely many
possible observables to chose in $\mathcal{B}(\mathcal{H}_{2})$.
Each one of these possibilities, defines a different
\emph{measurement context}. Thus, we can define the set
$$C_{A}=\{A\otimes B\,|\,B\in\mathcal{B}(\mathcal{H}_{2})\}$$
that enable us to represent all possible contexts associated to a
single observable $A\in\mathcal{B}(\mathcal{H}_{1})$. Now, prepare
$N$ copies of the compound system in the same state $\rho$ and
measure only observable $A$ of $S_{1}$. The state of $S_{1}$ is
given by $\rho_{1}=\mbox{tr}_{2}(\rho)$. From the preparation point
of view, all these processes are indistinguishable: we prepare $N$
copies of the compound system, and measure the \emph{same}
observable $A$ of $S_{1}$. But, even if these measurements are
indistinguishable for an observer focused only on $S_{1}$, they are
not exactly the same: a measurement of $A$ in $S_{1}$ could be
performed jointly with an arbitrary element of $C_{A}$, and these
elements could change on each run of the experiment. As an example,
if $N=N_{1}+N_{2}$, we can perform the experiment $A\otimes B$
$N_{1}$ times, and $N_{2}$ times the experiment $A\otimes B'$ (for
$B\neq B'$). Thus, each context defined on the compound system,
gives us a \emph{different} observable on the compound system, but
an indistinguishable one for system $S_{1}$. Thus, from the point of
view of the observer focused on subsystem $S_{1}$,  a measurement of
the property associated to the physical quantity $\mathcal{A}$ can
be represented by a collection of indistinguishable but yet
different observables.

There is a formalism that allows to deal properly with collections
of truly indiscernible entities, namely, quasi-set theory (for
details, see for example
\cite{krause_formal_1995,domenech_discussion_2007,domenech_q-spaces_2008,domenech_quantum_2010}).
In this theory, there is a primitive notion of indistinguishability,
represented by the symbol ``$\equiv$". The axioms are written in
such a way that, $X\equiv Y$ does not necessarily implies $X=Y$. The
equality symbol ``$=$'', can only be applied on special elements of
the theory (i.e., the classical ones). Thus, we can represent the
observable quantity $\mathcal{A}$ by a quasi-set, that we call
$[A]$, formed by the collection of all possible experimental
contexts associated to $\mathcal{A}$. The elements of $[A]$ are all
indistinguishable, but not the same (i.e., the quasi-cardinal
---which represents the number of elements--- of $[A]$ is greater
than $1$), in the sense that $X\equiv Y$, for all $X,Y\in [A]$. What
are the valuations compatible with this description? We must be
careful. Using quasi-set theory, if we try to assign a value to each
element of $[A]$, we define a quasi-function$f:[A]\longrightarrow
\{0,1,\cdots,d-1\}$ (here, we are assuming that $\mathcal{A}$
defines a $d$-dimensional observable). But, unlike the classical
case, were we have $d^{\sharp [A]}$ valuations, there are only $d$
quasi-functions of this kind, namely: $\langle [A],0\rangle,\langle
[A],1\rangle,\cdots,\langle [A],d-1\rangle$. This is so, due to the
fact that all ordered pairs collapse into the same class. This means
that, considered as an observable, the only values that $A$ can take
are given by $\{0,1,\cdots,d-1\}$. This is compatible with what we
observe in an actual experiment: the outcome set is given by
$\{0,1,\cdots,d-1\}$. But something more interesting happens when we
try to actually put values to the elements of $[A]$ previous to
measurement. Let us do this as follows. In order to pick up one
element of $[A]$, consider a strong singleton $[[X]]\subseteq [A]$.
In quasi-set notation, this means that $\mbox{qc}([[X]])=1$ and
$\forall x\in [[X]]$, we have $x\in [A]$. Now, take
$j\in\{0,1,\cdot,d-1\}$ and form the ordered pair
$\langle[[X]];j\rangle$ in such a way that the pair has only two
elements (this can be done in quasi-set theory). This pair can be
interpreted as follows: assign the value $j$ to the context $[[X]]$.
All contexts are indistinguishable from the perspective of the
observer associated to $S_{1}$, in the sense that, if the rest of
the universe is ignored, each representative of the class $[A]$ is
indistinguishable from the others. But they are different from the
point of view of the joint system, since the different contexts
define distinguishable global observables. By forming pairs of the
form $\langle[[X]];j\rangle$, a definite value can be assigned to an
observable in a given context. But an indistinguishable observable
taken from a different context may have a different value. In this
way, we see how a kind of logical indistinguishability, one taken
from quasi-set theory, can be used to model quantum contextuality in
a suitable way.

In order to illustrate the description of contextuality using
quasi-set theory with more detail, lets consider first a classical
system formed of three dichotomic random variables $X$, $Y$ and $Z$,
having values in the set $\{-1,1\}$. Let us assume that $X$, $Y$ and
$Z$ obey the classical theory of identity (and accordingly, that
these variables retain their identity among the different contexts
in which they may appear). Thus, for example, we are assuming that
$X$ is the same, independently of whether it is measured in
connection with $Y$ or in connection with $Z$. A similar
consideration holds for $Y$ and $Z$. In this way, we obtain the
following (classical) table:

\begin{table}[]
\begin{tabular}{|l|l|l|l|l|l|}
\hline
X & Y & Z & XY & XZ & YZ\\
\hline
1 & 1 & 1 & 1 & 1 & 1\\
1 & 1 & -1 & 1 & -1 & -1\\
1 & -1 & 1 & -1 & 1 & -1\\
-1 & 1 & 1 & -1 & -1 & 1\\
1 & -1 & -1 & -1 & -1 & 1\\
-1 & -1 & 1 & 1 & -1 & -1\\
-1 & 1 & -1 & -1 & 1 & -1\\
-1 & -1 & -1 & 1 & 1 & 1\\
\hline
\end{tabular}
\end{table}

Using the above table, a quick check indicates that the values the
compound random variables $XY$, $XZ$ and $YZ$, satisfy the
inequality
\begin{equation}
-1\leq XY+ XZ+ YZ\leq 3.
\end{equation}
Thus, by convexity, their mean values $\langle XY\rangle$, $\langle
XZ\rangle$ and $\langle YZ\rangle$ must, in turn, satisfy
\begin{equation}\label{e:ContextualityInequality}
-1\leq \langle XY\rangle+\langle XZ\rangle+\langle YZ\rangle\leq 3.
\end{equation}

Now, let us see what happens if, instead of assuming a classical
theory of identity, we consider that $X$, $Y$ and $Z$ define
indistinguishable properties. Thus, the only thing we can say is
that we have classes of indistinguishable properties $[X]$, $[Y]$
and $[Z]$, formed by all possible indistinguishables from $X$, $Y$
and $Z$, respectively. Thus, when considering, for example, $X$ in
connection with $Y$ and afterwards, in connection with $Z$, we will
have $X$ and $X'Z'$, with $X'$ indistinguishable from $X$ and $Z'$
indistinguishable from $Z$
---yet not the same! Thus, the value that we must assign to $X$
needs not to be the same than the one that we assign to $X'$.
Therefore, if we proceed as before and consider all possibilities,
we don't have a definite value for $X$, but a collection of them: a
value attached to each element of the class $[X]$, by appealing to
quasi-pairs of the form $\langle [[X]],-1\rangle$ and $\langle
[[X]],1\rangle$ (being $[[X]]\subseteq[X]$ a strong singleton of
$[X]$). A similar consideration holds for $Y$ and $Z$. The only
think we can do, regarding joint measurements, is to write down a
table like:

\begin{table}[]
\begin{tabular}{|l|l|l|l|l|l|}
\hline
X'Y' & X''Z' & Y''Z''\\
\hline
1 & 1 & 1\\
1 & -1 & -1\\
-1 & 1 & -1\\
-1 & -1 & 1\\
-1 & -1 & 1\\
1 & -1 & -1\\
-1 & 1 & -1\\
-1 & -1 & -1\\
1 & 1 & -1\\
1 & -1 & 1\\
-1 & 1 & 1\\
\hline
\end{tabular}
\end{table}

where the primed quantities are indistinguishables from $X$, $Y$ and
$Z$. Notice that the last four lines of the above table are strictly
forbidden for classical random variables. Or, in our terminology,
for random variables obeying the classical theory of identity. But
this implies that inequality \eqref{e:ContextualityInequality} will
be violated by random variables of this sort. Thus, we find that the
violation of the theory of identity for properties, can be used to
describe contextuality in a natural way.

It is important to remark that this is not a quantum example. The
reason is that, for three projection operators that commute
pairwise, it follows that it is possible to define a set of vectors
on  the Hilbert space such that those vectors are eigen-vectors of
those projectors. Therefore, it follows that a joint probability
distribution exists, and the properties associated to the projectors
are not contextual.  But, in principle, we could extend the argument
to a Bell-type scenario with four properties if necessary (though it
would be more cumbersome).

It is interesting to compare our approach with previous ones. The KS
theorem has a very straightforward consequence: to assume that a
given property possesses the same value among different contexts
leads to contradictions. This has led many authors to conclude
random variables --representing the same physical quantity-- are
different. In other words, that contexts can be used to index random
variables in such a way that they become different (see for example
\cite{dzhafarov_contextuality_2014}. According to the classical
theory of identity, this sounds as a reasonable conclusion. But
still, there remains a feeling that, with such a proliferation of
properties --in most examples of interest, there are, indeed,
infinitely many contexts for each given property-- a high
ontological price is paid, special when we consider that it is the
same physical quantity among all possible contexts. Our framework is
a way out of this situation, in the sense that, one is able to speak
about indistinguishable properties in a strong sense, but, at the
same time, the instances of these properties are not forced to
obtain the same values on each context in which they are considered.

\section{Conclusions}\label{sec:conclusions}

In this work we proposed a formal framework for dealing with
properties of contextual systems. According to our proposal,
properties describing the same physical quantity, but belonging to
different measurement contexts, are not different, but neither are
they equal. They are indistinguishable in a strong sense. The
existence of such objects require, mathematically, the introduction
of a theory that allows a violation of Leibnitz's principle of
identity of indiscernibles. Quasi-set theory is such a theory, a
set-theoretical framework that allows for the description of
collections of entities which do not obey the classical principle
identity
\cite{krause_quasi-set_1992,krause_formal_1995,krause_quasi-set_1999}.
Quasi-set theory includes objects that obey a weaker relationship of
indistinguishability, allowing for entities being different
\textit{solo numero}. Here we showed that quasi-sets can be used to
describe quantum contextuality in a consistent way. We believe this
approach opens the door for a new ontology for describing quantum
properties.

It is useful to compare our proposal with other ones. In the
contextuality by default approach (CdB), properties belonging to
different contexts are considered different \textit{ab initio}. If
this is done, we can assume that not only properties obey the
classical theory of identity, but they also satisfy classical
probability theory. The cost of such approach is that properties
then become context-dependent, and thus dependent on the observer's
choice of context. Our framework is an alternative one, that allows
to reconcile the fact that the same quantity may acquire different
values in different measurement contexts. In the other approaches
using non-standard probability, it is possible to argue (and we did
so in Section \ref{sec:indistinguishability-properties}) that the
use of indistinguishable objects from quasi-set leads to
non-standard probabilities. This, in a sense, may provides an
interpretation for the appearance of non-standard theories in
quantum physics, and we believe it is a topic that should be
investigated further. As such, it would also be interesting to study
the consequences for probability theory of assuming random variables
that do not obey the classical theory of identity. Such a
non-standard probability calculus may be useful to understand the
peculiar behavior of probabilities in quantum theory. In this
direction ---and, in connection with the problem of identical
particles--- we hope that by addressing contextuality in quantum
theory using quasi-sets, may help to understand the link between the
underlying quantum particle ontology and their properties.

The ontology behind our approach assumes that properties are, in
reality, associated to collections of indistinguishable entities.
This ontology has interesting consequences for answering the
following question: which are the necessary and sufficient
conditions for contextuality? According to our proposal,
contextuality appears whenever properties depart from classical
identity theory. But there could be many ways in which this may
happen, being quantum mechanics a particular case. This opens the
door for studying generalized probabilistic models using quasi-set
theory or similar set-theoretical frameworks. In particular, it
would be interesting to investigate which constraints in the
indistinguishability of properties would produce the quantum
boundaries, such as Tsirelson's.

The main advantage of our approach is that it is a more natural
description of what happens to properties in the quantum realm. The
main disadvantage is that it is not as general as other proposals,
such as contextuality by default, as it would not be applicable to
non-quantum contextual systems, since we expect such systems to not
have an ontological issue with identity. But, perhaps more
interesting from the point of view of this paper's authors, is that
our approach points toward a feasible quantum ontology ---perhaps a
physical principle--- that needs to be thought in more detail.

\vskip6pt

\enlargethispage{20pt}


\section*{Acknowledgments}

The authors would like to thank Pawel Kurzinsky for discussions on
the issue of contextuality and indistinguishability. We also
benefited from conversations with participants at the Winer Memorial
Lectures held at Purdue University in November 2018.

\bibliographystyle{plain}
\bibliography{references}
\end{document}